\documentclass[review,3p,11pt,a4paper,times]{elsarticle}

\usepackage{graphicx}
\graphicspath{{Figures/}} 
\usepackage{amssymb}
\usepackage{mathtools}
\usepackage{amsthm}
\usepackage{xcolor}

\usepackage{lineno}
\usepackage{amsmath}

\usepackage[short,c1]{optidef}
\usepackage{booktabs,subcaption,amsfonts,dcolumn}
\usepackage{times}

\setcitestyle{authoryear,open={(},close={)}}

\journal{XXXX}

\begin{document}

\begin{frontmatter}


\title{A Flexible Rolling Regression Framework for Time-Varying SIRD models: Application to COVID-19}



\author[Javier]{Javier Rubio-Herrero
\corref{cor1}}
\ead{javier.rubioherrero@unt.edu}
\cortext[cor1]{Corresponding author}
\author[Edwin]{Yuchen Wang}
\ead{Yuchen.Wang@unt.edu} 

\address{Department of Information Technology and Decision Sciences, G. Brint Ryan College of Business, University of North Texas, 1155 Union Circle, Denton, TX 76201}

\begin{abstract}
The present paper introduces a data-driven framework for describing the time-varying nature of an SIRD model in the context of COVID-19. By embedding a rolling regression in a mixed integer bilevel nonlinear programming problem, our aim is to provide the research community with a model that reproduces accurately the observed changes in the number of infected, recovered, and death cases, while providing information about the time dependency of the parameters that govern the SIRD model. We propose this optimization model and a genetic algorithm to tackle its solution. Moreover, we test this algorithm with 2020 COVID-19 data from the state of Minnesota and found that our results are consistent both qualitatively and quantitatively, thus proving that the framework proposed is an effective an flexible tool to describe the dynamics of a pandemic.
\end{abstract}


\end{frontmatter}



\section{Introduction}

In December 2019, an outbreak of viral pneumonia of unknown origin occurred in Wuhan, Hubei Province, China. A new type of coronavirus by the name of \textit{2019 Novel Coronavirus} (COVID-19) was isolated from the confirmed patients. COVID-19, a single-stranded RNA virus of the genus $\beta$, is characterized by a long incubation period and high contagiousness. Common symptoms after infection include cough, fever, diarrhea, and difficulty breathing \citep{capano2020mobilizing}.

Shortly after the outbreak in China, COVID-19 spread throughout the whole world and became a pandemic. 
By the end of 2020, the total number of confirmed cases of infections reached 100 million globally, whereas 1.8 million people worldwide had died as a consequence of this pandemic, including 350,000 in the United States \citep{hopkins}. It is well known that the main transmission route of COVID-19 is by droplets and close contact, but it is also possible that it can be contracted via aerosol transmission. Amidst the fight against this pandemic, various countries and regions have introduced policies to curb its spread. Measures such as confinements, stay-at-home orders, remote work, face masks, as well as other policies for controlling and preventing further spread within and among communities, have become common.

COVID-19 is the second pandemic of the XXI century after the H1N1 pandemic of 2009. After the \textit{Spanish ﬂu} that in 1918 caused between 20 and 40 million casualties all over the world \citep{trilla20081918,world2005avian}, the current pandemic has been the most severe in the last century and a half. Since its outbreak, countries worldwide have faced shortages in testing capacity, lack of effective drugs, and insufficient knowledge to handle and control COVID-19, especially during the early stages of the pandemic. As a consequence, countries have struggled seriously with keeping their healthcare systems and facilities under maximum capacity. Moreover, COVID-19 has also taken a heavy toll on the global economy. For example, in the United States, the unemployment rate soared temporarily to $14.8\%$ shortly after the outbreak \citep{unempl_USA}, the highest since the \textit{Great Depression} that started in 1929.

As in other pandemics, the battle against COVID is being fought on several fronts: immunization, treatment, isolation, and training \citep{laarabi2013optimal}. Due to progress in the clinical understanding of the disease and the increasing experience gained in treating this virus, mortality rates had dropped by the time 2020 came to an end. Also, as this manuscript is being written, immunization efforts have already started worldwide. While there is a common goal in keeping the spread of this contagious virus at bay, vaccination policies vary amongst countries, albeit it is the norm that especial priority is given to healthcare workers, older people, and to those that, because of preexisting conditions, fall within a group of risk \citep{world2005influenza}.

\section{Background on mathematical modeling in epidemiology}\label{sec:backgrounds}

Mathematical models have lied historically in the heart of epidemiology analysis. \cite{kermack1927contribution} studied the epidemics of \textit{Black Death} in London and the plague in Mumbai in 1906 using a revolutionary \textit{susceptible-infected-removed} (SIR) model that identified population dynamics with the evolution of these epidemics:

\begin{eqnarray*}
\frac{dS(t)}{dt}&=&-\beta \frac{S(t) I(t)}{N},\\
\frac{dI(t)}{dt}&=&\beta \frac{S(t)I(t)}{N} - \gamma I(t),\\
\frac{dR(t)}{dt}&=&\gamma I(t),
\end{eqnarray*}

Where $S(t),I(t),R(t)$ indicate the number of individuals that are susceptible, infected, and recovered at time $t$, respectively, and $N$ is the total population under study. The parameter $\beta\ge0$ represents the average number of contacts of an infected individual per unit of time, multiplied by the probability of disease transmission. $\gamma\ge0$ is the proportion of infected individual that recover from the disease (per unit of time). With these models
 came the well-known \textit{threshold theory}, supported by the definition of a \textit{basic reproduction number}, which aims to describe whether a disease is prevalent or receding.

Since this seminal work, countless research efforts have been made in this area, very often with highly accurate descriptive and predictive results. Thus, it does not come as a surprise that these models have been used for forecasting the likelihood and severity of the COVID-19 pandemic. Examples of recent works are  \cite{wangping2020extended,kucharski2020early,yang2020modified,fanelli2020analysis,xue2020data,postnikov2020estimation,li2020modeling} and \cite{cooper2020sir}, to name a few.

In general,  well-known extensions of the original SIR model can be consulted in \cite{hethcote1989three,hethcote2000mathematics,hethcote2009basic} and \cite{weiss2013sir}. While a detailed description of the many known variations of the basic SIR model is out of the scope of this paper, we will mention several that have been proposed in the context of COVID-19. For example, \cite{giordano2020modelling} further expanded the SIR model to  a SIDARTHE model considering diagnosed, ailing, and dead cases, and found the prediction rates fitted the actual rates after lockdown or social distancing order issued by local government. Even models with more compartments have been used \citep{ndairou2020mathematical}. However, as commented in \cite{roda2020difficult}, more complicated models do not necessarily yield better results. As a matter of fact, \cite{ifguis2020simulation} predicted accurately COVID-19 cases in Morocco with a simulation based on an SIR model. \cite{croccolo2020spreading}, on the other hand, presented a percolation-type model based on an SIR model that aimed at mimicking the effect of lockdowns in the United States. In any case, most epidemiology models of this sort are complicated to solve analytically and it is common to resort to simulation or other techniques to attain solutions(e.g., the work by \cite{khoshnaw2017identifying} with \textit{Elzaki transforms}).



 The model's parameters play an essential role in characterizing the evolution of a disease in the framework of SIR or SIR-related models. For this reason, their study has brought the attention of many researchers. In many instances, these parameters are hypothesized based on previous experience or assumptions. This is the case in \cite{simha2020simple}, where a standard SIR model is modified with a Brownian motion to demonstrate the randomness of the virus spread. On other occasions, the parameters are inferred from available data. For example,  \cite{postnikov2020estimation} fitted the parameters of a simple SIR model using actual data to demonstrate that these models are applicable in this context (very much in line with what \cite{roda2020difficult} suggested). 
  
The basic SIR model introduced above uses constant parameters or rates. However, there are time-dependent epidemiology models that allow their rates to be a function of time. This enhances these models' flexibility, but it also introduces additional complexity in estimating these rates. Despite this hindrance, these time-dependent models are commonly used. There exist analytic methods by which it is possible to obtain time series that represent the model parameters. However, these methods rely on assumptions that might not always hold. For example, \cite{zhong2020early}, and \cite{waqas2020analysis} assume that the number of infected people is negligible concerning the number of susceptible individuals. While this is reasonable during the early stages of a pandemic, it can be questioned later. 
  
Most of the data-driven approaches that aim at inferring the model's parameters resort to some fitting method. Such is the case of \cite{yang2016fitting}, who built a time-dependent SIR model of the Dengue virus to fit coefficients of transmission over four different time periods. In the context of COVID-19, this was also done by \cite{teles2020time} when the authors divided the time series in periods of 5 days according to the virus incubation period and fitted the parameters of their model accordingly. In \cite{godio2020seir}, the authors used a metaheuristic approach to fit the parameters of a \textit{susceptible-exposed-infected-recovered} (SEIR) model to predict the evolution of the current pandemic in Italy and compare it with Spain and South Korea. Some other works assume that the model parameters can be characterized by \textit{finite impulse responses} (FIRs) and employ \textit{ridge regression} to find the coefficients of these responses \citep{chen2020time} or can be found by using a linear combination of basis functions whose parameters are fitted with sparse identification techniques \citep{calafiore2020modified}. Black-box models like \textit{deep neural networks} have also found their application in this field and can be used to find how these parameters change over time, although inherent to the use of these methods is the fact that closed-form equations are not attained \citep{jo2020analysis}.
  
Rolling regression is a time-series modeling technique that is often used in finance and economics \citep{fabozzi2011equity,dunis2004applied, zanin2012rolling}. It finds applicability in contexts where regression coefficients fluctuate over time and substitute a static regression approach for a dynamic method. This increases the flexibility of the regression model and improves its capacity for prediction. Rolling regression requires selecting a window size that is rolled over time in steps of, typically, one time period. This window size determines how many data points are selected for evaluating the regression coefficients. When rolled, the regression analysis is updated, and so are the corresponding coefficients. Despite its convenience for time-series analysis, rolling regression has not received much attention in epidemiology, in general, and in the study of the current pandemic, in particular. An exception to which we will frequently refer throughout this paper is \cite{anastassopoulou2020data}. However, we believe that the highly-contagious COVID-19 virus and its shifts in latency and symptoms appearance for different groups make rolling regression an appealing alternative for modeling purposes.
 
  
The rolling regression used in \cite{anastassopoulou2020data} is performed with a window of 6 days and, like \cite{zhong2020early} and \cite{waqas2020analysis}, they assume that the total population can approximate the number of susceptible individuals. This assumption facilitates the analysis in the early stages of the pandemic but seems insufficient for characterizing its dynamics later. Also, \cite{anastassopoulou2020data} rightfully acknowledges that the observed infected and recovered individuals are just a fraction of the actual ones and carry their study under two separate and predefined scenarios; these fractions vary. \cite{calafiore2020modified} assume that these fractions are equal to each other and other fraction of deceased individuals and calculate this value by gridding.
  
Inspired by the models presented by \cite{calafiore2020modified}, and \cite{anastassopoulou2020data}, the present paper introduces an approach that could be considered a ``hybrid" of these two. On the one hand, we propose a time-varying model where we estimate the parameters of a \textit{susceptible-infected-recovered-deceased} (SIRD) model via rolling regression with a window size that is not fixed beforehand; on the other hand, our SIRD model also considers that the observed infected, recovered, and deceased individuals are fractions of the actual numbers, but we do not assume that these fractions have the same value. Instead, we opt for an optimization framework in which the data in hand determine these parameters (i.e., fractions and window size). Thus, this paper addresses a descriptive effort whose goals we summarize below:

  \begin{enumerate}
      \item To provide researchers with a general, data-driven, and flexible framework for modeling the dynamics of this pandemic. This framework is constituted by a \textit{mixed integer bilevel nonlinear programming} (MIBNLP) model in which a rolling regression is the \textit{lower-level} (or \textit{follower}) optimization problem. This MIBNLP derives empirically (i.e., based on the data available) the proportions of the actual number of individuals in each state (S, I, R, D) that our data represent and the optimal window size such that the SIRD model fits these data in the most accurate manner. We will approach the solution of this problem with a \textit{genetic algorithm}. 
      \item We test our algorithm to solve the proposed MIBNLP with 2020 data from the state of Minnesota, in the United States. The optimization model will yield time series of the parameters of the SIRD model. We will analyze these results both qualitatively and quantitatively. The qualitative assessment will explain the evolution of the obtained time-varying parameters of the SIRD model given the government policies in this state. The quantitative assessment will study the sensitivity of our results against a parameter of the MIBNLP model that controls the smoothness of the resulting time series. We will also compare the summary statistics of these time series with previously published works.
    \end{enumerate}
  
 The rest of this paper is organized as follows: in Section \ref{sec:RR_framework} we introduce our selected epidemiology model, the optimization problem in which the rolling regression is embedded, and we introduce our solution approach; next, we discuss our results in Section \ref{sec:Results}; finally, we wrap up this work with our conclusions in Section \ref{sec:Conclusions}.

\section{Rolling regression framework for obtaining time-dependent parameters of the SIRD model}\label{sec:RR_framework}
\subsection{The epidemiology model}

Following \cite{calafiore2020modified}, we carry our analysis with a time-dependent \textit{susceptible-infected-recovered-deceased} (SIRD) model:

\begin{eqnarray}
    \frac{dS(t)}{dt}&=&-\beta(t) \frac{S(t)I(t)}{(S(t)+I(t))},\label{eqn:susceptible}\\
    \frac{dI(t)}{dt}&=&\beta(t) \frac{S(t)I(t)}{(S(t)+I(t))}-\gamma(t) I(t) - \mu(t) I(t),\label{eqn:infected}\\
    \frac{dR(t)}{dt}&=&\gamma(t) I(t),\label{eqn:recovered}\\
    \frac{dD(t)}{dt}&=&\mu(t) I(t),\label{eqn:deceased}
\end{eqnarray}
\noindent where $S(t),I(t),R(t),D(t)$ denote the number of susceptible, infected, recovered, and deceased individuals over time, respectively. The time-varying parameters $\beta(t),\gamma(t)\ge0$ have the same meaning described for the classic SIR model (see Section \ref{sec:backgrounds}) and $\mu(t)\ge0$ is the proportion (per unit time) that an infected individual dies as a consequence of the disease. Let us define the vector of variables $\mathbf{y}(t) \doteq [S(t),I(t),R(t),D(t)]^T$ and its time-derivative $\mathbf{\dot{y}}(t)\doteq d\mathbf{y}(t)/dt$. Since we disregard the effect of births and deaths due to causes other than COVID-19, it is implicit in the formulation above that $\mathbf{\dot{y}}(t)=\mathbf{0},t=1,\dots,T,$ and thus we assume that the population of individuals, $N$, remains constant throughout the time of study. For a given number of infected, recovered, and deceased people at time $t$, the number of susceptible individuals can be calculated straightforwardly as $S(t) = N-I(t)-R(t)-D(t)$. Therefore, when the states and their change over time are known, the time-dependent parameters $\beta(t),\gamma(t),\mu(t)$ can be estimated empirically with equations (\ref{eqn:susceptible})-(\ref{eqn:deceased}). If we define the vector of parameters $\boldsymbol{\theta}(t)\doteq[\beta(t),\gamma(t),\mu(t)]^T$, these equations can be re-written in matrix form as a system of linear equations, $\mathbf{\dot{y}}(t)=\mathbf{A}(t)\boldsymbol{\theta}(t)$, where

\begin{equation*}
    \mathbf{A}(t) = 
    \begin{bmatrix}
    -\frac{S(t)I(t)}{(S(t)+I(t))} & 0 & 0\\
    \frac{S(t)I(t)}{(S(t)+I(t))} & -I(t) & -I(t)\\
    0 & I(t) & 0\\
    0 & 0 & I(t)
    \end{bmatrix}.\label{eqn:SIRD_matrix}
\end{equation*}

However, in reality, the actual number of infected, recovered, and deceased individuals is not accurately known. Therefore, using directly the \textit{observed} individuals in each state in the SIRD model will result in important inaccuracies when estimating the rates of the model. \cite{chen2020time} or \cite{anastassopoulou2020data} already addressed this issue in the context of COVID-19 (the former with an analytical approach; the latter in a more empirical fashion). In this work, we propose a generalization of the solution proposed by \cite{calafiore2020modified}, who assumed that the data collected represents a fraction of the actual data. However, these authors assume that this proportion is the same for all states, namely, $S(t), I(t), R(t)$ and $D(t)$. Our approach suggests different proportions that need to be estimated based on the data available.

Let us define $\alpha_X,X\doteq I,R,D$ as the proportions of the actual individuals in each compartment that are observed by our data, with $0\le\alpha_X\le1$. Then the observed stocks can be defined as $\tilde{X}=\alpha_X X$ Further, we define $\tilde{S}\doteq\alpha_{S}S$, with $0\le\alpha_{S}\le1$.The four proportions can be included in a vector $\boldsymbol{\nu}\doteq[\alpha_{S},\alpha_{I},\alpha_{R},\alpha_{D}]^T$. Hence, letting $\mathbf{\tilde{y}}(t) \doteq [\tilde{S}(t),\tilde{I}(t),\tilde{R}(t),\tilde{D}(t)]^T=$ and $\mathbf{\dot{\tilde{y}}}(t)\doteq d\mathbf{\tilde{y}}(t)/dt$, our linear system of equations can be reformulated as $\mathbf{\dot{\tilde{y}}}(t)=\mathbf{\tilde{A}}(t,\boldsymbol{\nu})\boldsymbol{\theta}(t)$, where

\begin{equation*}
    \mathbf{\tilde{A}}(t,\boldsymbol{\nu})=
    \begin{bmatrix}
    -\alpha_{S}\frac{\tilde{S}(t)\tilde{I}(t)}{(\alpha_I\tilde{S}(t)+\alpha_S\tilde{I}(t))} & 0 & 0\\
    \alpha_{I}\frac{\tilde{S}(t)\tilde{I}(t)}{(\alpha_I\tilde{S}(t)+\alpha_S\tilde{I}(t))} & -\tilde{I}(t) & -\tilde{I}(t)\\
    0 & \frac{\alpha_R}{\alpha_I}\tilde{I}(t) & 0\\
    0 & 0 & \frac{\alpha_D}{\alpha_I}\tilde{I}(t)
    \end{bmatrix}. 
\end{equation*}

Note that given values for $\alpha_{S},\alpha_{I},\alpha_{R},$ and $\alpha_{D}$, we can calculate $\tilde{S}(t)=\alpha_{S}\left(N-\tilde{I}(t)/\alpha_{I}-\tilde{R}(t)/\alpha_{R}-\tilde{D}(t)/\alpha_{D}\right)$.  The definition and usage of $\alpha_S$ is merely formal and it lacks meaning, since there is not such as thing as \textit{observed} susceptible stock versus \textit{actual} susceptible stock. However, its addition will act as a ``normalizer" that will reduce the order of magnitude of $S(t)$ and help in the fitting process that will be introduced shortly. We will discuss this further in Section \ref{sec:Results}. Since the data available were collected daily (i.e., at discrete intervals of time), $\mathbf{\dot{\tilde{y}}}(t)$ can be expressed in the form of variations or changes and thus we will work with the following system of equations:
\begin{equation}
\mathbf{\Delta\tilde{y}}(t)=\mathbf{\tilde{A}}(t,\boldsymbol{\nu})\boldsymbol{\theta}(t),\label{eqn:discrete_SIRD}
\end{equation}
\noindent where $\Delta(\cdot)=(\cdot)(t)-(\cdot)(t-1)$.

\subsection{The rolling regression framework}
We propose a rolling regression framework that solves Equation (\ref{eqn:discrete_SIRD}) for the vector $\boldsymbol{\theta}(t)$  in the least-squares sense by applying multivariate multiple regression. The rolling regression takes place with a window size $w$ and a step of one unit of time (i.e., one day). If we have data for the period $t=1,\dots,T$, the result of this approach will be a series of vectors $\boldsymbol{\hat{\theta}}(t),t=w,w+1,\dots,T$. The proposed framework is characterized for its flexibility since we look for the window size and the vector of proportions $\boldsymbol{\nu}$ that provide the best fit for a given set of data. We will work with an optimization problem that can be cast as follows:

    \begin{mini!}[0]
        {\boldsymbol{\nu},w}{\frac{1}{T-w}\sum_{t=w}^T \|\boldsymbol{\Delta\hat{\tilde{y}}}(t,\boldsymbol{\nu})-\boldsymbol{\Delta\tilde{y}}(t)\|_2+\lambda\sum_{t=w+1}^T\|\boldsymbol{\Delta\hat{\theta}}(t,\boldsymbol{\nu})\|_2}{\label{eqn:RR_obj}}{}
        {\label{eqn:RR_optimizer}}
        \addConstraint{\boldsymbol{\Delta\hat{\tilde{y}}}(t,\boldsymbol{\nu})}{=\mathbf{\tilde{A}}(t,\boldsymbol{\nu})\boldsymbol{\hat{\theta}}(t,\boldsymbol{\nu}),\quad\label{eqn:RR_estimation}}{t=w,\dots,T}
        \addConstraint{\mathbf{0}\le}{\boldsymbol{\nu}\le\mathbf{1}\label{eqn:RR_nu_var}}{}
        \addConstraint{w_{min}\le}{w\le w_{max}}{}
        \addConstraint{w}{\in\mathbb{Z^+}\label{eqn:RR_w_integer}}{}
        \addConstraint{\boldsymbol{\hat{\theta}}(t,\boldsymbol{\nu})}{  =\arg\min_{\boldsymbol{\theta}\ge\mathbf{0}}\sum_{\tau=t-w}^t \|\mathbf{\tilde{A}}(\tau,\boldsymbol{\nu})\boldsymbol{\theta}-\boldsymbol{\Delta\tilde{y}}(\tau)\|_2^2,\quad\label{eqn:RR_multivariate}}{t=w,\dots,T}
    \end{mini!}

The optimization problem defined by equations (\ref{eqn:RR_optimizer})-(\ref{eqn:RR_multivariate}) is a \textit{mixed integer bilevel nonlinear problem} (MIBNLP) with $T-w$ lower-level problems corresponding to all the multivariate multiple regressions that conform our rolling regression over the time of study. It aims at minimizing the average difference between the variations in the compartments as calculated with the least-squares time series of parameters $\boldsymbol{\hat{\theta}}(t,\boldsymbol{\nu})$, $\boldsymbol{\Delta\hat{\tilde{y}}}(t,\boldsymbol{\nu})$, and the variations in the states as given by our data, $\boldsymbol{\Delta\tilde{y}}(t)$.  The calculation of the vectors $\boldsymbol{\Delta\hat{\tilde{y}}}(t,\boldsymbol{\nu})$ takes place in constraint (\ref{eqn:RR_estimation}) which, in turn, uses the vectors $\boldsymbol{\hat{\theta}}(t,\boldsymbol{\nu})$ that solve the multivariate multiple linear regression problem specified in constraint (\ref{eqn:RR_multivariate}). Note that the parameters $\boldsymbol{\hat{\theta}}(t,\boldsymbol{\nu})$ fitted for the window $\{t-w,t\}$ are used to calculate $\boldsymbol{\Delta\hat{\tilde{y}}}(t,\boldsymbol{\nu})$. Hence, there are $T-w$ vectors $\boldsymbol{\Delta\hat{\tilde{y}}}(t,\boldsymbol{\nu})$ (i.e., $T-w$ vectors $\boldsymbol{\hat{\theta}}(t,\boldsymbol{\nu})$) and $T$ vectors $\boldsymbol{\Delta\tilde{y}}(t)$.

This optimization model looks for the optimal vector of proportions and the optimal window size to perform those regressions. The latter is set to be an integer between two bounds defined by the modeler, $w_{min}$ and $w_{max}$. In this application, we expect $w_{min}$ to be a small integer (we will see that in our case, we set $w_{min}=5$) and, in these circumstances, the first term in Equation (\ref{eqn:RR_optimizer}) will most likely find smaller errors in smaller windows. This results in optimal values of $w$ trivially equal to $w_{min}$. The downside is that it also results in very noisy time series for $\boldsymbol{\hat{\theta}}(t,\boldsymbol{\nu})$, as these parameters will be calculated based on just a few observations. Therefore, they are very sensitive to the updates that take place when the window is rolled. To avoid this, the objective function (\ref{eqn:RR_optimizer}) incorporates a second term that penalizes sudden increments in $\boldsymbol{\hat{\theta}}(t,\boldsymbol{\nu})$ with a penalty factor $\lambda$. It is the modeler's decision to tune this parameters in order to find a compromise between the quality of the fit and the smoothness of the time series defined by the vectors $\boldsymbol{\hat{\theta}}(t,\boldsymbol{\nu})$.

\subsection{Solution approach}

\begin{figure}[h!]
\centering
\includegraphics[scale=0.15]{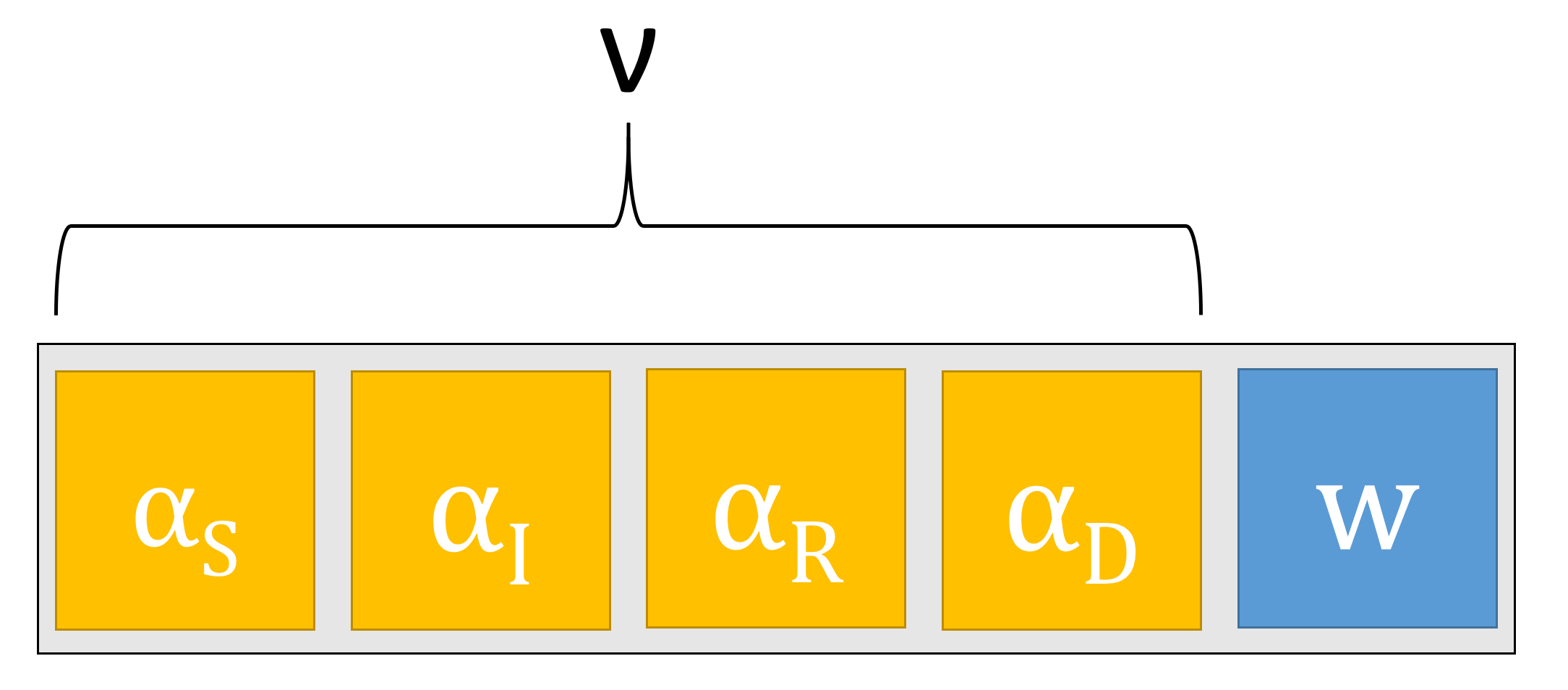}
\caption{Structure of an individual in the genetic algorithm}
\label{fig:GA_individual}
\end{figure}

In order to solve the MIBNLP above we opted for a real-valued \textit{genetic algorithm} \citep{dreo2006metaheuristics,haupt2004practical}. The individuals in our metaheuristic are vectors $[\boldsymbol{\nu},w]$ (see Figure \ref{fig:GA_individual}). For a given individual that is feasible for the constraints (\ref{eqn:RR_nu_var})-(\ref{eqn:RR_w_integer}), there are $T-w$ multivariate multiple regressions that can be run (corresponding to the $T-w$ lower-level optimization problems), to obtain the time series of the parameters of the SIRD model. Then, a series of $T-w$ vectors of predicted variations can be calculated directly according to constraint (\ref{eqn:RR_estimation}).

\begin{figure}[h!]
\centering
\includegraphics[scale=0.45]{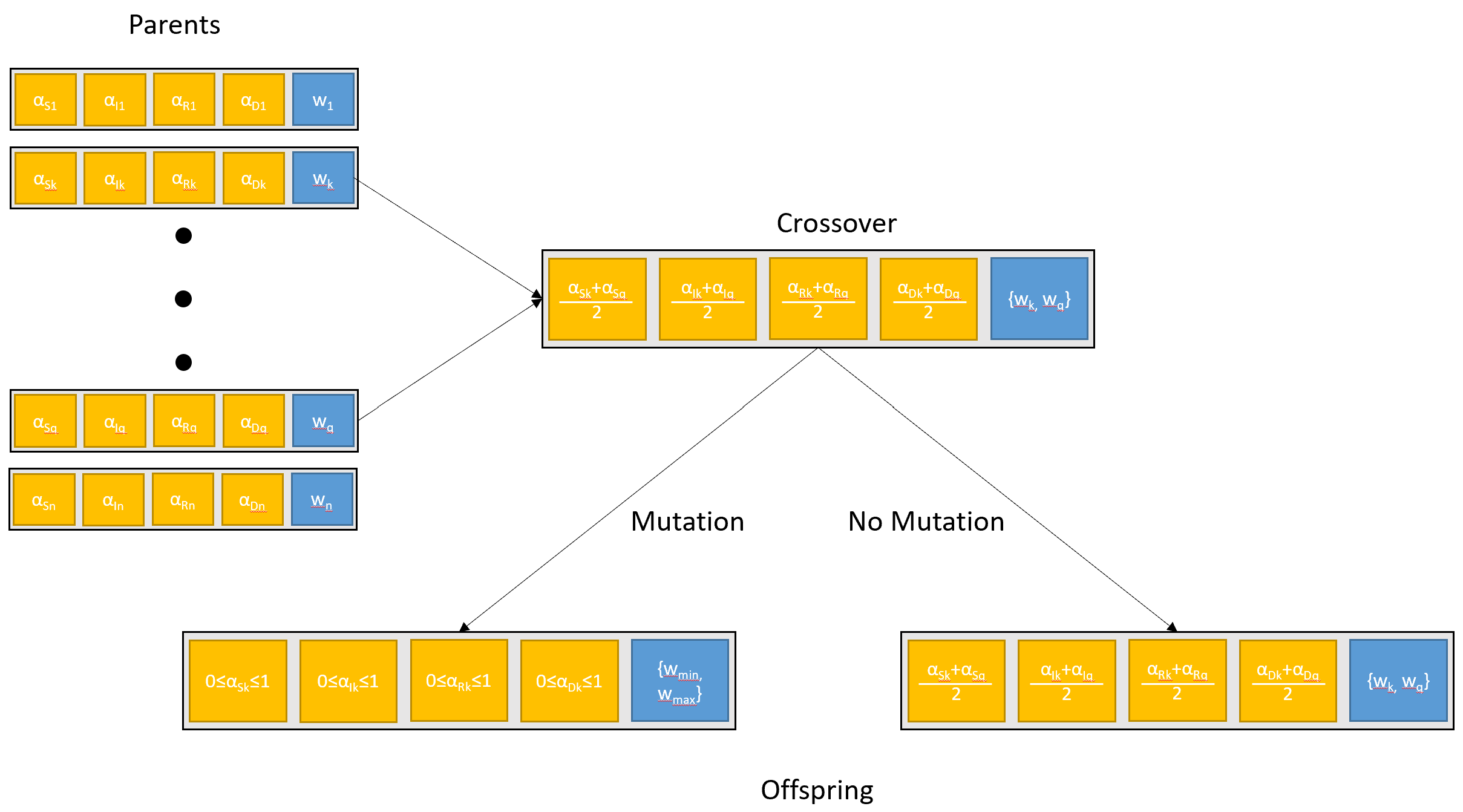}
\caption{Generation of individuals in the genetic algorithm}
\label{fig:GA_process}
\end{figure}

How modelers set the parameters of a genetic algorithm depends largely on their experience and the application in hand. Aiming at finding an equilibrium between population diversity and time for convergence to a solution, we followed the recommendation given in \cite{storn1997differential} and set the initial population size to $10$ times the number of elements in an individual (i.e., we selected a size of $50$ for our first generation of individuals). For the crossover stage, individuals were selected randomly in pairs. When it comes to the elements in $\boldsymbol{\nu}$, for each pair of individuals crossed over, the resulting offspring was generated with the average of each parent's values. For the value of $w$, an integer value between the parents' values of $w$ was drawn randomly according to a discrete uniform distribution. With $50$ individuals in each generation, we produced an offspring of size $25$, where each child was then allowed to mutate in the range of $[0,1]$ for each element of $\boldsymbol{\nu}$ and in the set $\{w_{min},w_{min}+1,\cdots,w_{max}\}$ for $w$. After attempting different mutation rates, an acceptable equilibrium between the heterogeneity of the population and convergence was obtained with a probability of mutation of $0.2$. From the pool of parents and children, the best $50$ advanced to the next generation. Therefore, we followed a \textit{best-in-class} selection that guaranteed that the fitness function's value was non-decreasing as generations passed. This fitness function was chosen to be Equation (\ref{eqn:RR_optimizer}). After a minimum of $40$ generations, if for $10$ consecutive generations, the difference in the fitness of the best individuals of two consecutive generations was smaller than a threshold $\epsilon=10^{-6}$, then the genetic algorithm was halted. Abusing of notation, the best solution thus far was declared as ``optimal", even though it is clear that such claim is highly unlikely when working with metaheuristics. A graphical representation of this algorithm is presented in Figure \ref{fig:GA_process}.         

\section{Results}\label{sec:Results}
\subsection{Quantitative Analysis of the Results}

\begin{figure}[h!]
    \centering
    \includegraphics[scale=0.17]{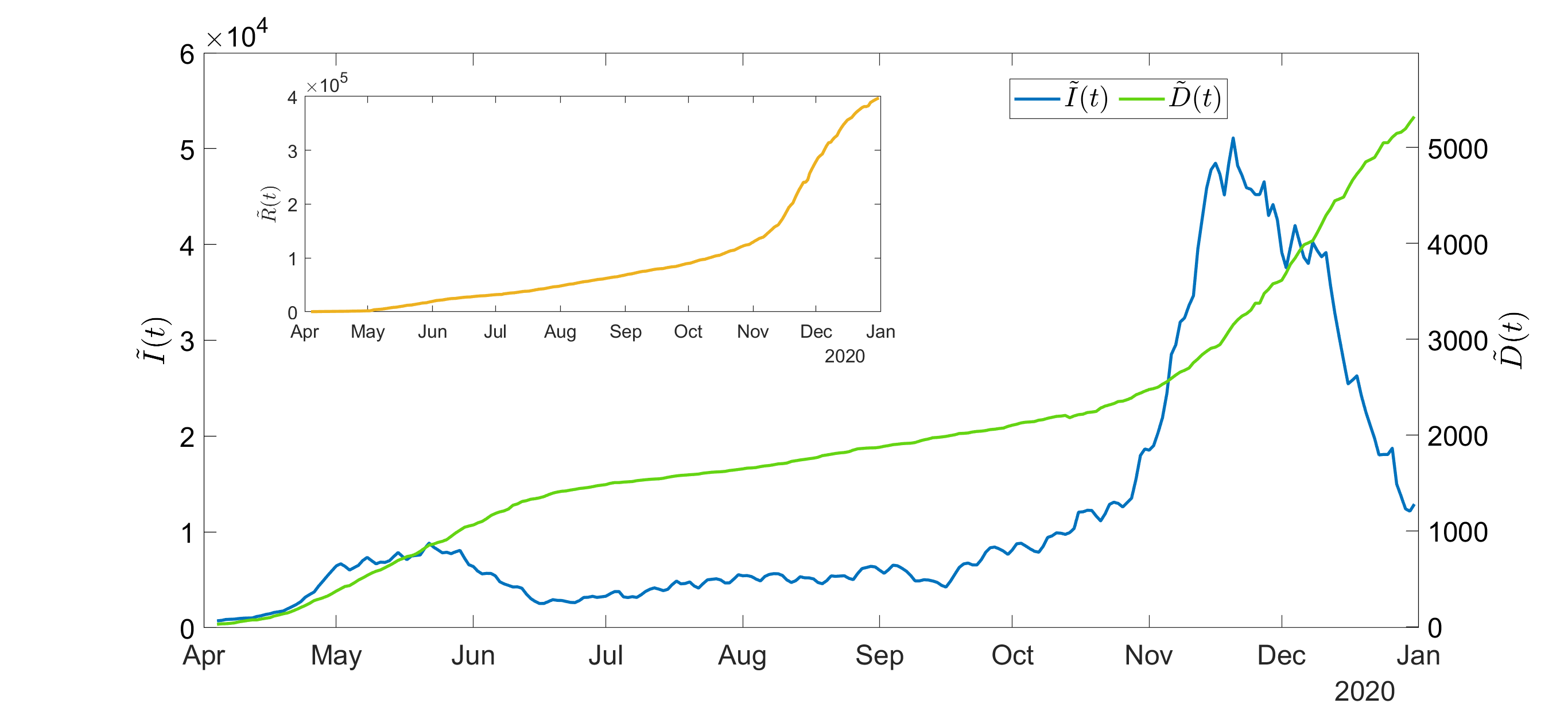}
    \caption{5-day average of the number of infected, recovered, and dead individuals as consequence of COVID-19 in Minnesota}
    \label{fig:data_MN}
\end{figure}

The framework proposed by the optimization problem (\ref{eqn:RR_optimizer})-(\ref{eqn:RR_multivariate}) was tested with COVID-19 data from Minnesota. These data, which covered the period between April and December of 2020, consisted of daily numbers on infected, recovered, and dead individuals and was retrieved from \textit{The COVID Tracking Project} \citep{COVID_tracking} (see Figure \ref{fig:data_MN}). The time series collected was smoothed with a 5-day average to dampen the usual noise introduced when new data are entered long after the number of infected, recovered, and dead people have changed. This happens especially on Mondays when databases usually record changes that occurred during the weekend.

We solved several instances of the MIBNLP problem as to assess the impact of the penalty parameter $\lambda$ on the model's solution. Starting with $\lambda=0$, we increased its value until the weight of the penalty term was sufficiently important to steer the optimal value of $w$, $w^*$, away from $w_{min}$. This lower bound was set to $w_{min}=5$. This was similar to the window of $6$ days used by \cite{anastassopoulou2020data}. We did not set $w_{min}$ to a lower value in order to avoid trivial regressions, but we allowed for a sufficiently wide range of options by setting $w_{max}=30$.

\begin{figure}[h!]
    \centering
    \begin{subfigure}[h!]{1\textwidth}
        \centering
        \includegraphics[scale=0.17]{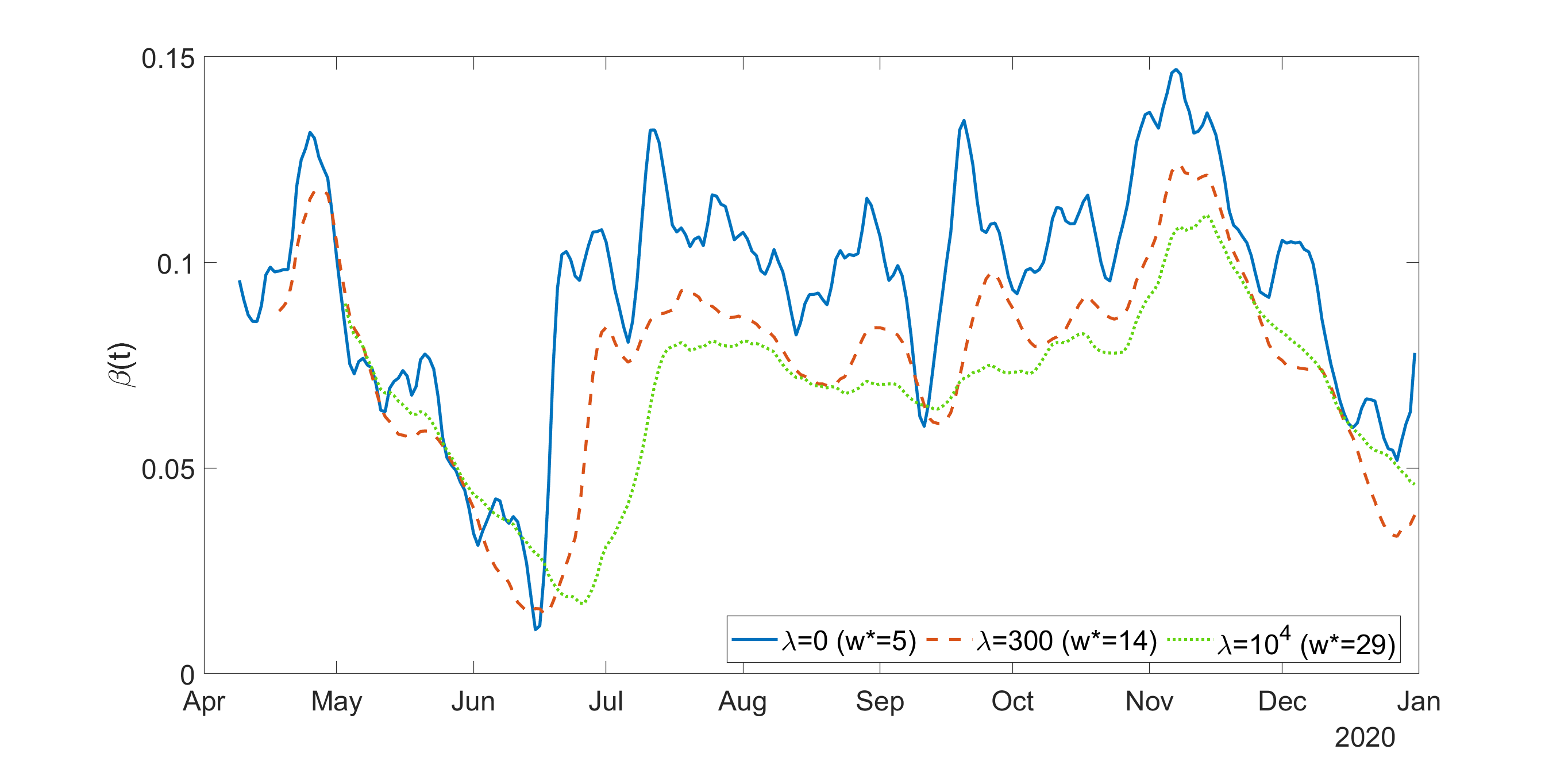}
    \end{subfigure}
    \begin{subfigure}[h!]{1\textwidth}
        \centering
        \includegraphics[scale=0.17]{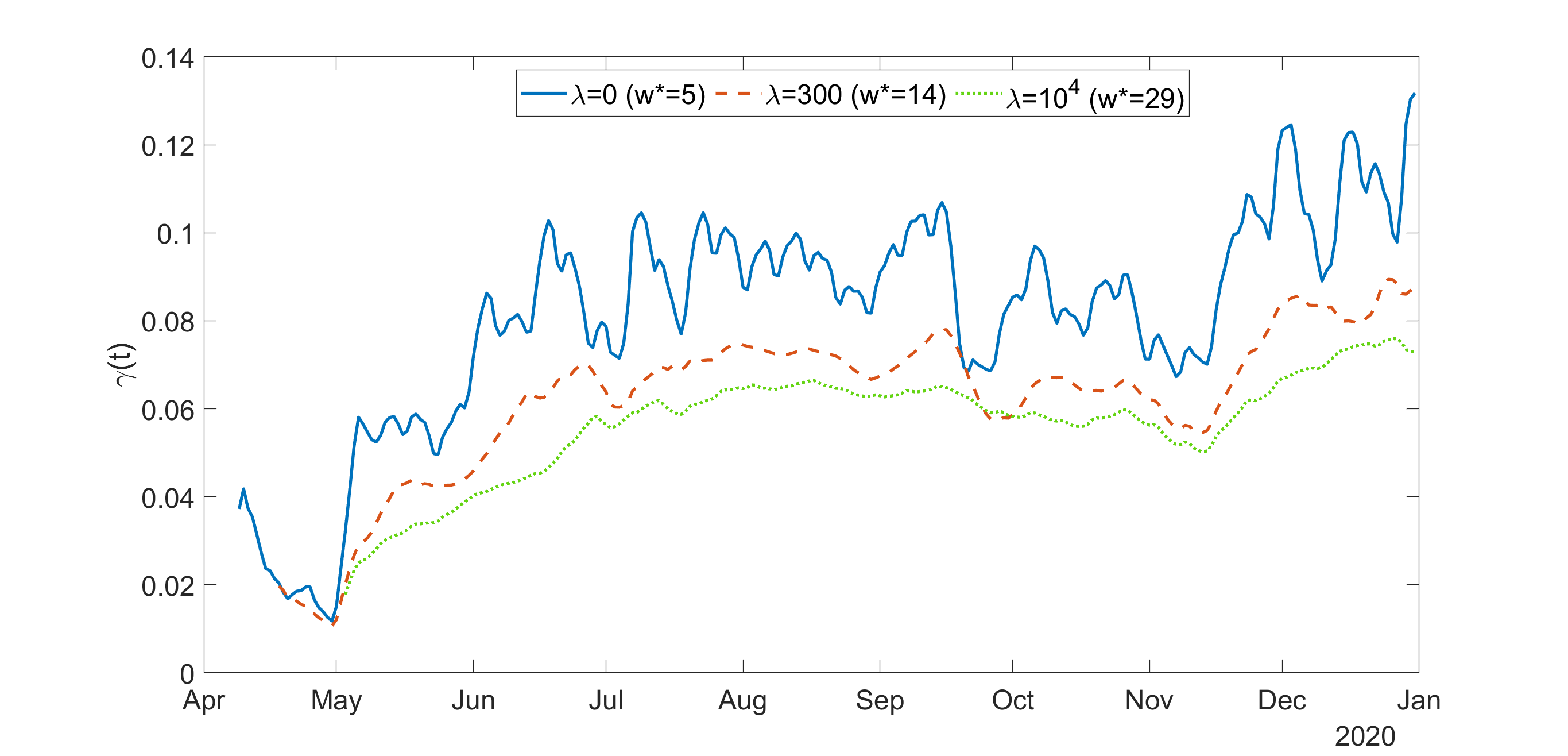}
    \end{subfigure}
    \begin{subfigure}[h!]{1\textwidth}
        \centering
        \includegraphics[scale=0.17]{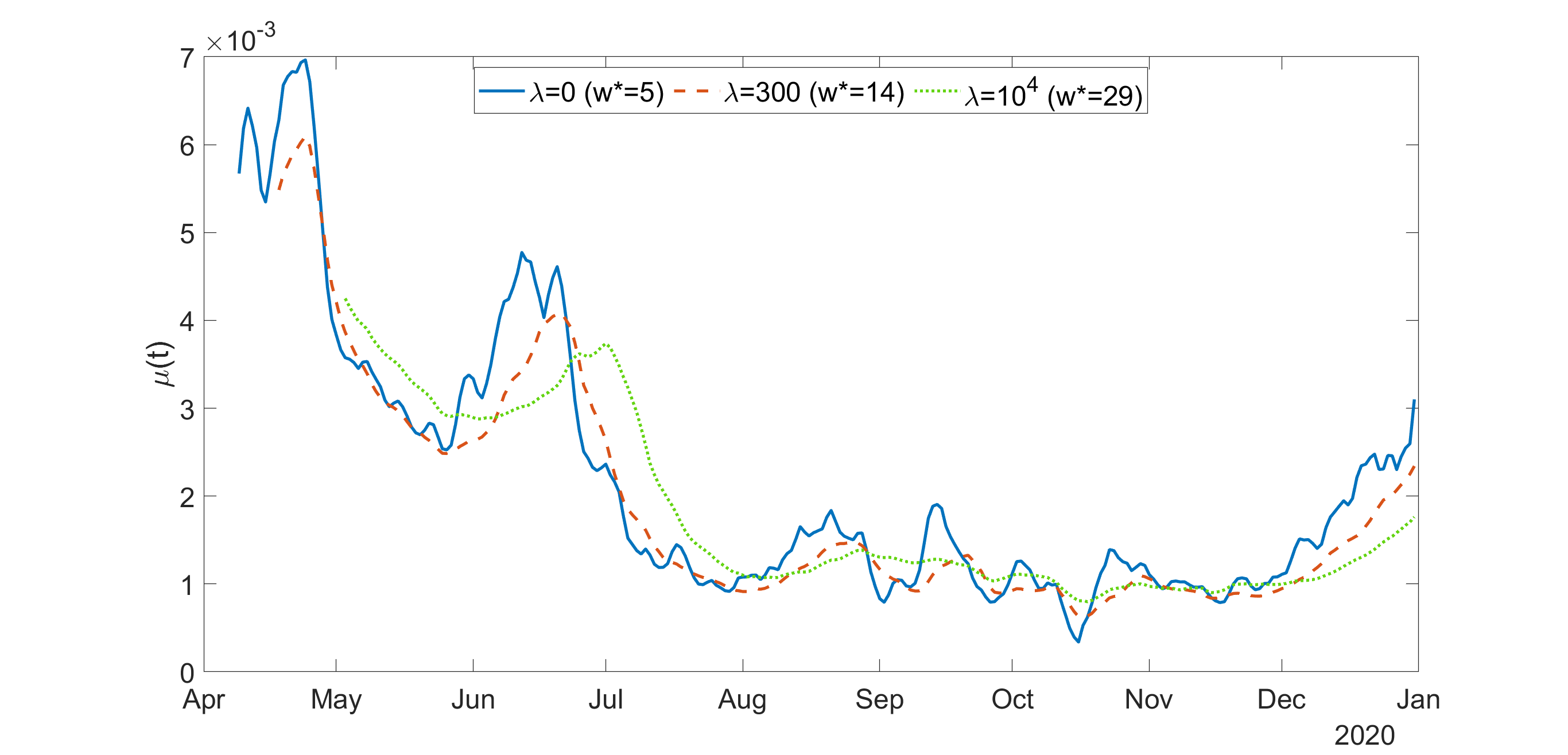}
    \end{subfigure}
    \caption{Effect of different penalty parameters on the fitted time series of parameters}
    \label{fig:lambda_chart_MN}
\end{figure}

When testing for different values of the penalty parameter, we found that the time-series of $\beta(t),\gamma(t),\nu(t)$ were smoother as $\lambda$ increased. This behavior was expected, as the penalty term in (\ref{eqn:RR_optimizer}) increases with $\lambda$. Smoother functions are attained when windows are larger because the changes incurred in the data used for the regression every time the window is rolled tend to have a smaller impact as the sample size increases. The results are shown in Figure \ref{fig:lambda_chart_MN} where, the higher $\lambda$, the higher the optimal window size. Note that different window sizes produce curves with different time lags.

\begin{figure}[h!]
    \centering
    \begin{subfigure}[h!]{1\textwidth}
        \centering
        \includegraphics[scale=0.13]{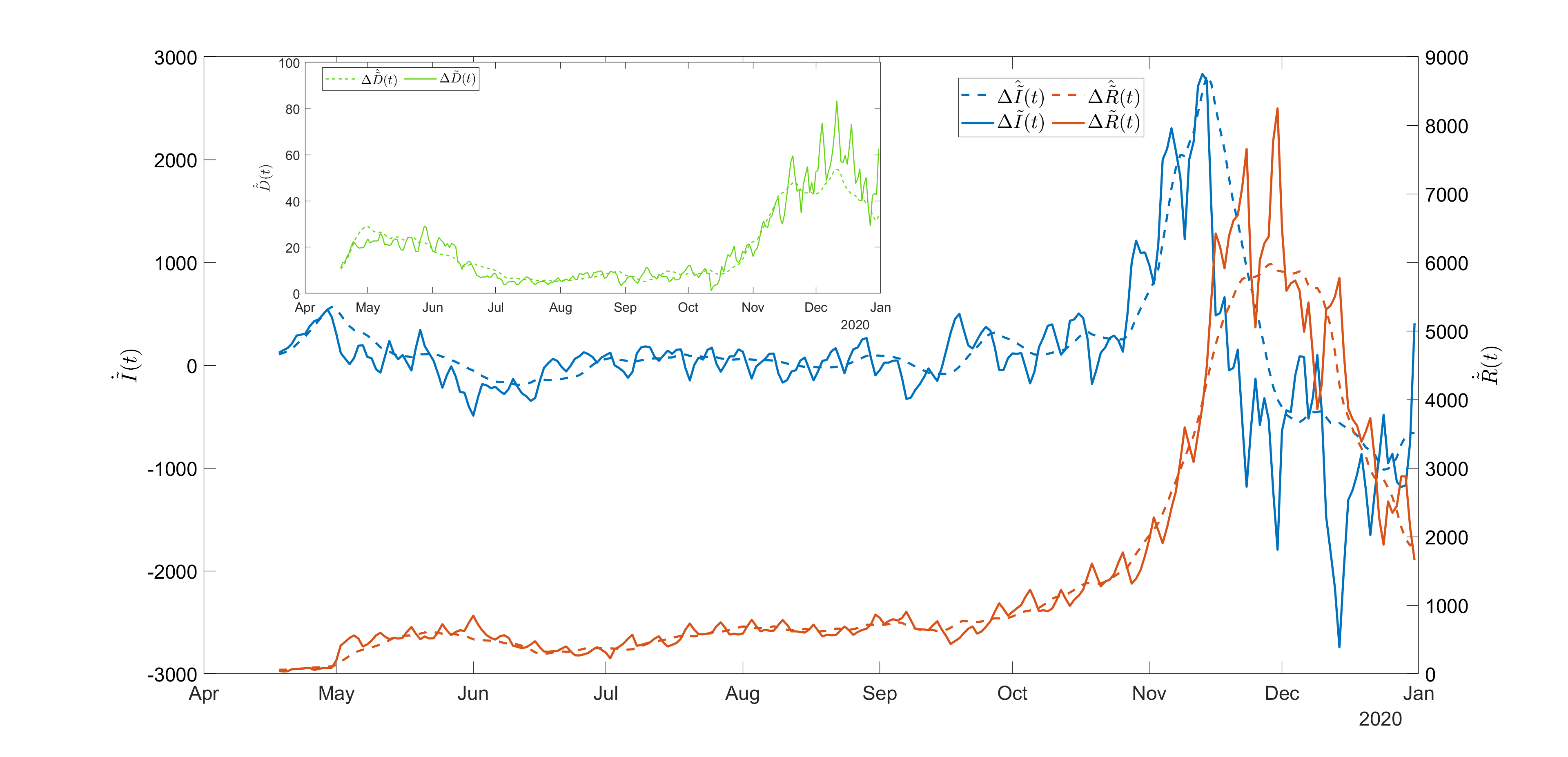}
    \end{subfigure}
    \caption{Predicted variations of the number of infected, recovered, and deceased individuals ($\lambda=300$)}
    \label{fig:variations_lambda}
\end{figure}

The time series of $\beta(t),\gamma(t)$, and $\mu(t)$ that resulted from solving the optimization model (\ref{eqn:RR_optimizer})-(\ref{eqn:RR_multivariate}) (i.e., $\boldsymbol{\hat{\theta}}(t,\boldsymbol{\nu}^*)$) were substituted in Equation (\ref{eqn:RR_estimation}),
\begin{eqnarray*}
    \boldsymbol{\Delta\hat{\tilde{y}}}(t,\boldsymbol{\nu^*})&=&\mathbf{\tilde{A}}(t,\boldsymbol{\nu^*})\boldsymbol{\hat{\theta}}(t,\boldsymbol{\nu^*}),\quad t=w,\dots,T,
\end{eqnarray*}
\noindent and compared with our actual time series of state variations, $\boldsymbol{\Delta\tilde{y}}(t)$. Figure \ref{fig:variations_lambda} shows our results for a moderate value of $\lambda$ ($\lambda=300$). By \textit{moderate} we mean a value that yielded an optimal window size that was not very close to neither $w_{min}$ nor $w_{max}$. We show that the best solutions obtained for $\lambda=300$ produced time series of variations of the states of the model that followed the data quite accurately. We observe better predictions until the second half of November, followed by period of less accuracy characterized by sudden changes in the data. Note that the variations in the number of recovered and deceased people are always non-negative. This is because both states are terminal states, meaning that the stock in these states in non-decreasing over time in an SIRD model. The infected stock, however, may see periods of increase and decrease, as it is a state of transition.

\begin{figure}[h!]
    \centering
    \begin{subfigure}[h!]{1\textwidth}
        \centering
        \includegraphics[scale=0.17]{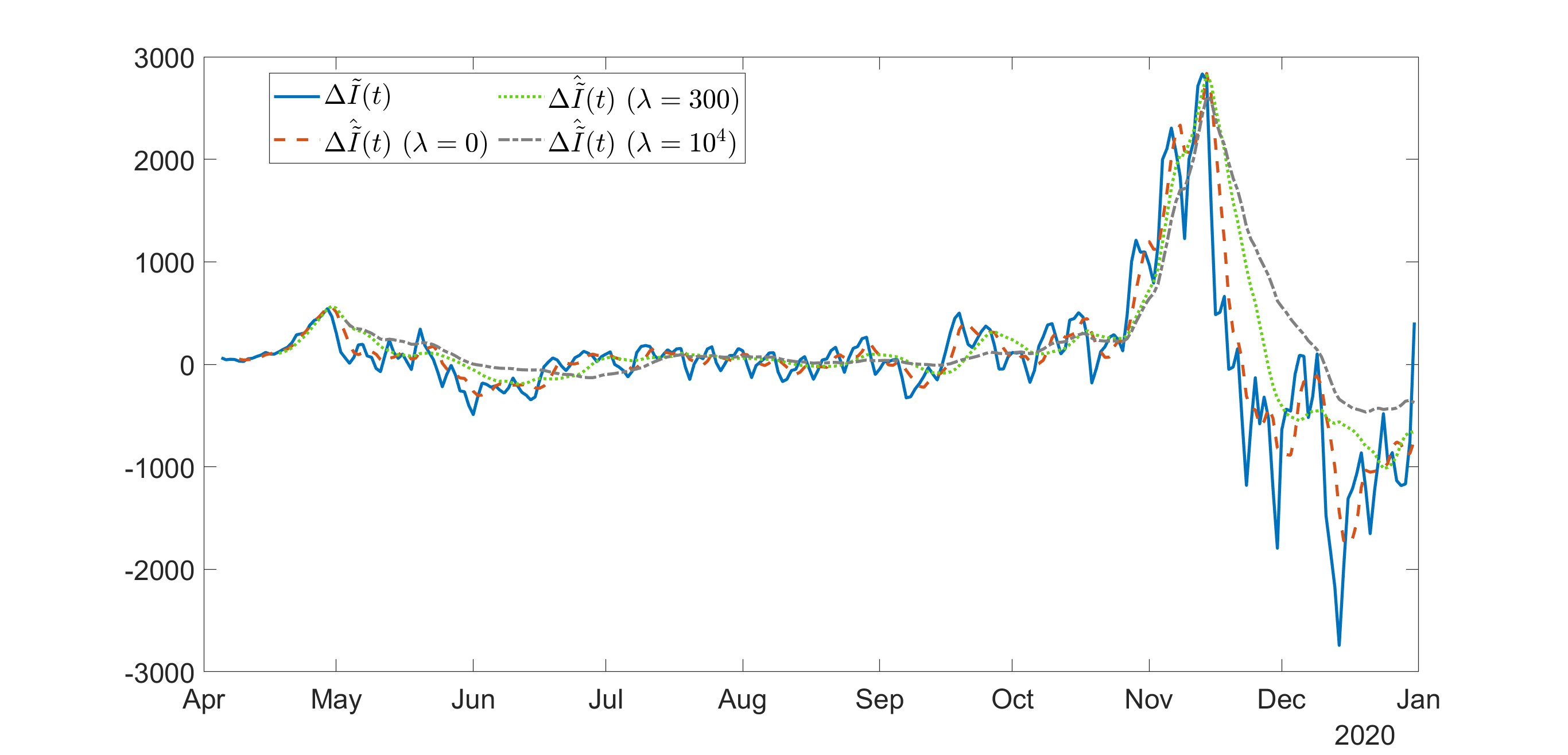}
        \caption{Infected}
        \label{fig:infected_lambda}
    \end{subfigure}
        \begin{subfigure}[h!]{1\textwidth}
        \centering
        \includegraphics[scale=0.17]{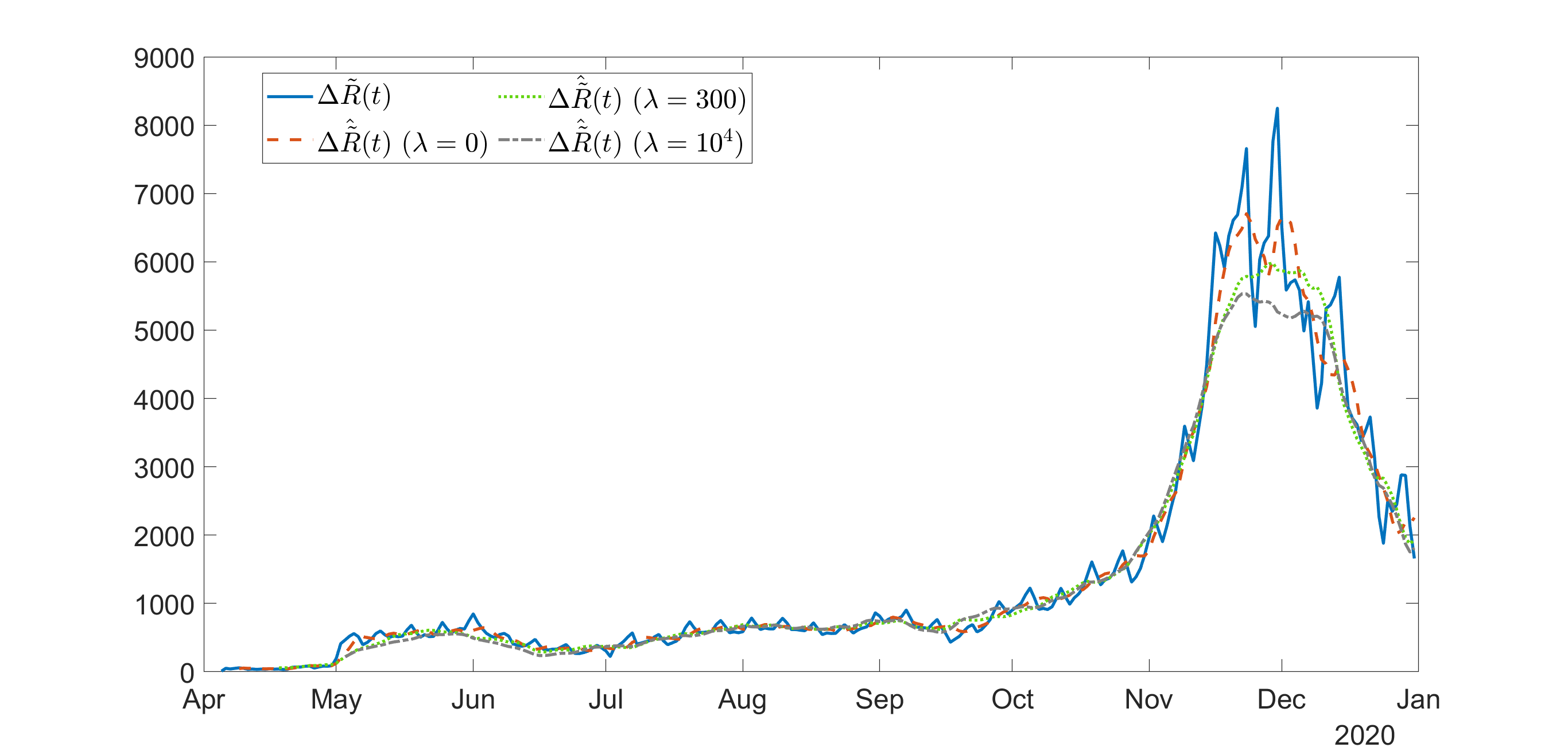}
        \caption{Recovered}
        \label{fig:recovered_lambda}
    \end{subfigure}
        \begin{subfigure}[h!]{1\textwidth}
        \centering
        \includegraphics[scale=0.17]{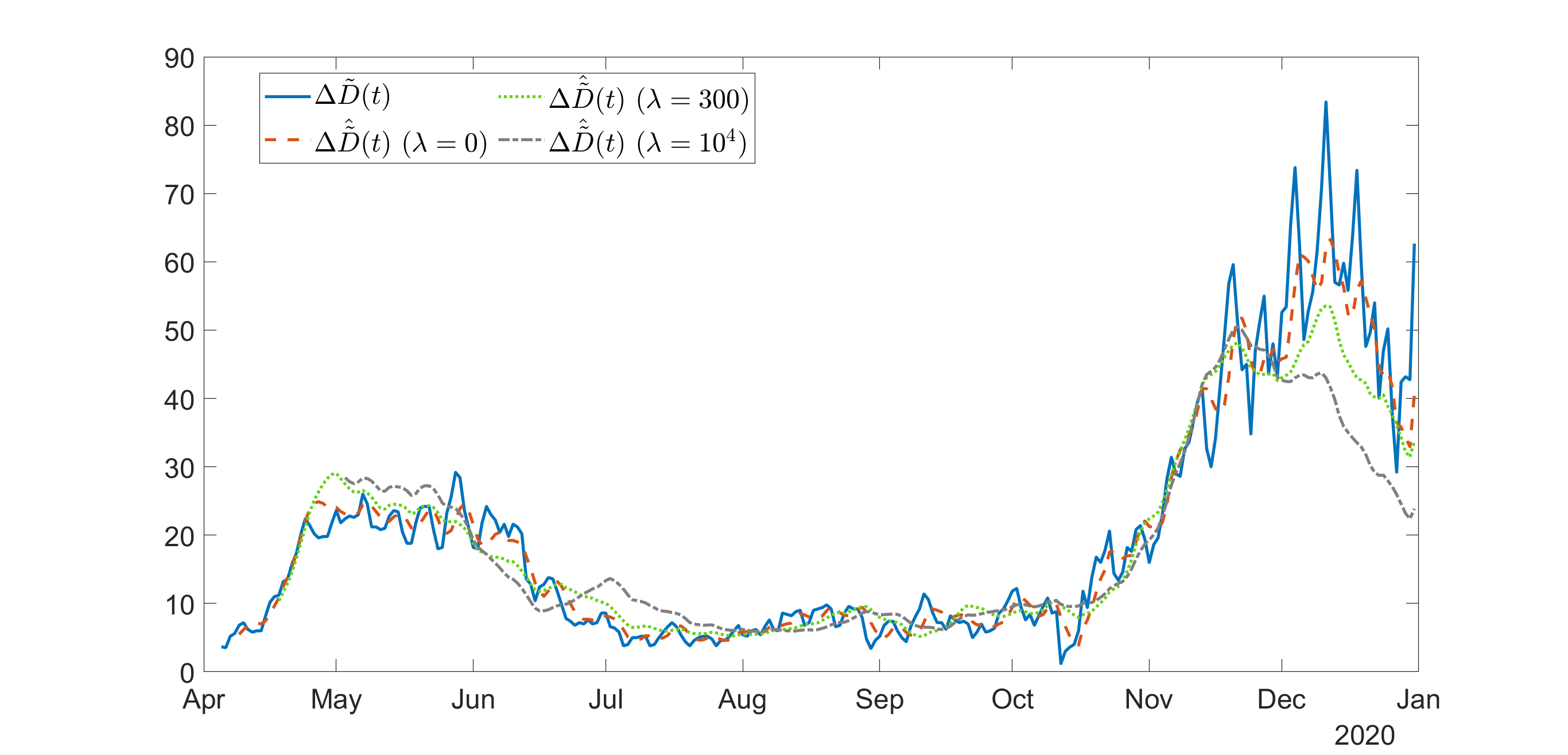}
        \caption{Dead}
        \label{fig:dead_lambda}
    \end{subfigure}
    \caption{Accuracy of predictions as a function of $\lambda$}
    \label{fig:predicted_lambda}
\end{figure}

Next, we assess the trade-off between $\lambda$ and the accuracy of the predictions. This is shown in Figure \ref{fig:predicted_lambda}. Small values of $\lambda$ allow reduce the smoothness of the predicted variations considerably in order to adapt to the noisy data. As $\lambda$ increases, the difference between predictions and data is more pronounced at the end of the year, when higher variations occurred after Thanksgiving. Given that the predictions are being made over noisy information, it seems sensible that a moderate value of $\lambda$ will result in a more accurate representation of the reality.  

On the other hand, Table \ref{tab:summary_stats} summarizes the descriptive statistics of the time series found with moderate values of $\lambda$. The order of magnitude of the parameters found is in agreement with other epidemiological studies. For example,  \cite{calafiore2020modified} analyzed the parameters in two different stages (\textit{early} versus \textit{late} stages of the pandemic) and attained values of $\beta(t)$ and $\gamma(t)$ in the order of $\backsim 10^{-2}$ and in the order of $\backsim 10^{-3}$ for $\mu(t)$ for the late stage. With our time series starting in May 2020, a comparison against this particular stage is fair. These orders of magnitude are average for Minnesota. \cite{chakraborty2020analyzing} reports values of $\beta(t)$ and $\mu(t)$ in the United States of orders $\backsim 10^{-2}$ and $\backsim 10^{-3}$, respectively, around the middle of May. This is also consistent with the average values of $\boldsymbol{\beta}(t)$ and slightly higher than minimum value of $\boldsymbol{\mu}(t)$ for Minnesota at that time of the year, as can be inferred from the variations of infected and recovered individuals in Figure \ref{fig:variations_lambda}.

\begin{table}[h!]
 \begin{subtable}[t]{1\textwidth}
\centering
\footnotesize{
\begin{tabular}{ |l|c|c|c| }
 \toprule
  Date Range & \multicolumn{3}{|c|}{04/18/2020 - 12/31/2020}\\
\hline
 Penalty factor ($\lambda$)&   \multicolumn{3}{|c|}{$300$}\\
 \hline
 & $\boldsymbol{\beta}(t)$ (d\textsuperscript{-1}) & $\boldsymbol{\gamma}(t)$ (d\textsuperscript{-1})& $\boldsymbol{\mu}(t)$ (d\textsuperscript{-1})\\
 \midrule
 Min    &$1.46\cdot 10^{-2}$&$1.07\cdot 10^{-2}$&$6.27\cdot 10^{-4}$\\
 Max    &   $1.24\cdot 10^{-1}$&$8.26\cdot 10^{-2}$&$6.09\cdot 10^{-3}$\\
 Mean &$7.79\cdot 10^{-2}$&$5.95\cdot 10^{-2}$&$1.92\cdot 10^{-3}$\\
 Median & $8.31\cdot 10^{-2}$&$6.49\cdot 10^{-2}$&$1.22\cdot 10^{-3}$\\
 Std. Dev.& $2.51\cdot 10^{-2}$&$1.65\cdot 10^{-2}$&$1.34\cdot 10^{-3}$\\
\bottomrule
\end{tabular}
 }
 \end{subtable}
 \caption{Summary statistics of the optimal time series found for two states of the United States}
 \label{tab:summary_stats}
\end{table}

 The best parameters obtained with the genetic algorithm and what their values imply are also an important discussion. These are shown in Table \ref{tab:optimal_param} for three different values of $\lambda$.
 The disparity in the values of the proportions suggests that the our framework benefits greatly from the flexibility introduced by allowing different proportions. Our results for $\alpha_I^*$ are in line with the \textit{Centers for Disease Control and Preventions's} (CDC) belief that the actual number of infected people in the United States was two to seven times greater than the reported cases by the end of the 2020 (i.e., $0.1429\le\alpha_I\ge 0.5$). According to a recent study \citep{noh2021estimation}, the values of $\alpha_I^*$ obtained for Minnesota are within the 95\% CI of the ascertainment of infected cases in this state until September 3, 2020. Assessing the accuracy of $\alpha_R^*,$ and $\alpha_D^*$ seems complicated due to testing issues and the variability in testing and recording practices among different parts of the United States. The actual death toll due to COVID-19 is unknown and very often researchers have to study the increase in deaths due to related conditions in order to make an estimation.   For example, \cite{weinberger2020estimating} use the excess in deaths from pneumonia and influenza (P\&I) to estimate that the death toll due to COVID-19 in New Jersey can be two or three times the official count (i.e., $0.33\le\alpha_D\ge 0.5$). Except for the result obtained for $\lambda=300$ ($\alpha_D^*=0.6171$), which is slightly more optimistic about this estimation,  our results for Minnesota are also within this range. 
 
 Finally, we comment on the best values found for $\alpha_S^*$. This parameter does not have a real meaning because there is not a distinction between observed and actual susceptible stock. However, its introduction is useful for regression purposes as it helps reduce the number of susceptible individuals introduced in the model, which is, by far, the largest stock in the model. With such low values of $\alpha_S^*$, we have that $\tilde{S}(t)\ll S(t)$ and therefore we reduce the order of magnitude of the errors incurred by our regression model when predicting the susceptible stock. Using $S(t)$ instead would have focused our model's regression efforts in predicting accurately the changes in this variable, several orders of magnitude larger than the rest, thus disregarding important prediction errors in $\tilde{I}(t),\tilde{R}(t)$, and $\tilde{D}(t)$. The small values of $\alpha_S^*$ obtained in all the instances that were run with our genetic algorithm indicate that the algorithm always seeks the ``normalization" of $S(t)$ in order to find good fits for the data provided.

\begin{table}[h!]
\footnotesize{
\centering
\begin{tabular}[t]{|c|c|c|c|c|c| }
 \toprule
 $\lambda$&$\alpha_S^*$& $\alpha_I^*$ (d\textsuperscript{-1}) & $\alpha_R^*$ (d\textsuperscript{-1})& $\alpha_D^*$ (d\textsuperscript{-1}) & $w^*$ (day)\\
 \midrule
$0$&$0.0167$ & $0.5215$  & $0.6921$&$0.5298$ & $5$\\
$\mathbf{300}$&$\mathbf{0.0556}$ & $\mathbf{0.3570}$  & $\mathbf{0.6349}$&$\mathbf{0.6171}$ & $\mathbf{14}$\\
$10^4$&$0.0749$ & $0.4476$  & $0.8613$&$0.4703$ & $29$\\
\bottomrule
\end{tabular}
 \caption{Best individuals found with the genetic algorithm for different values of $\lambda$}
 \label{tab:optimal_param}
 }
 \end{table}

\subsection{Qualitative Analysis of the Results}

\begin{figure}[h!]
\centering
\includegraphics[scale=0.4]{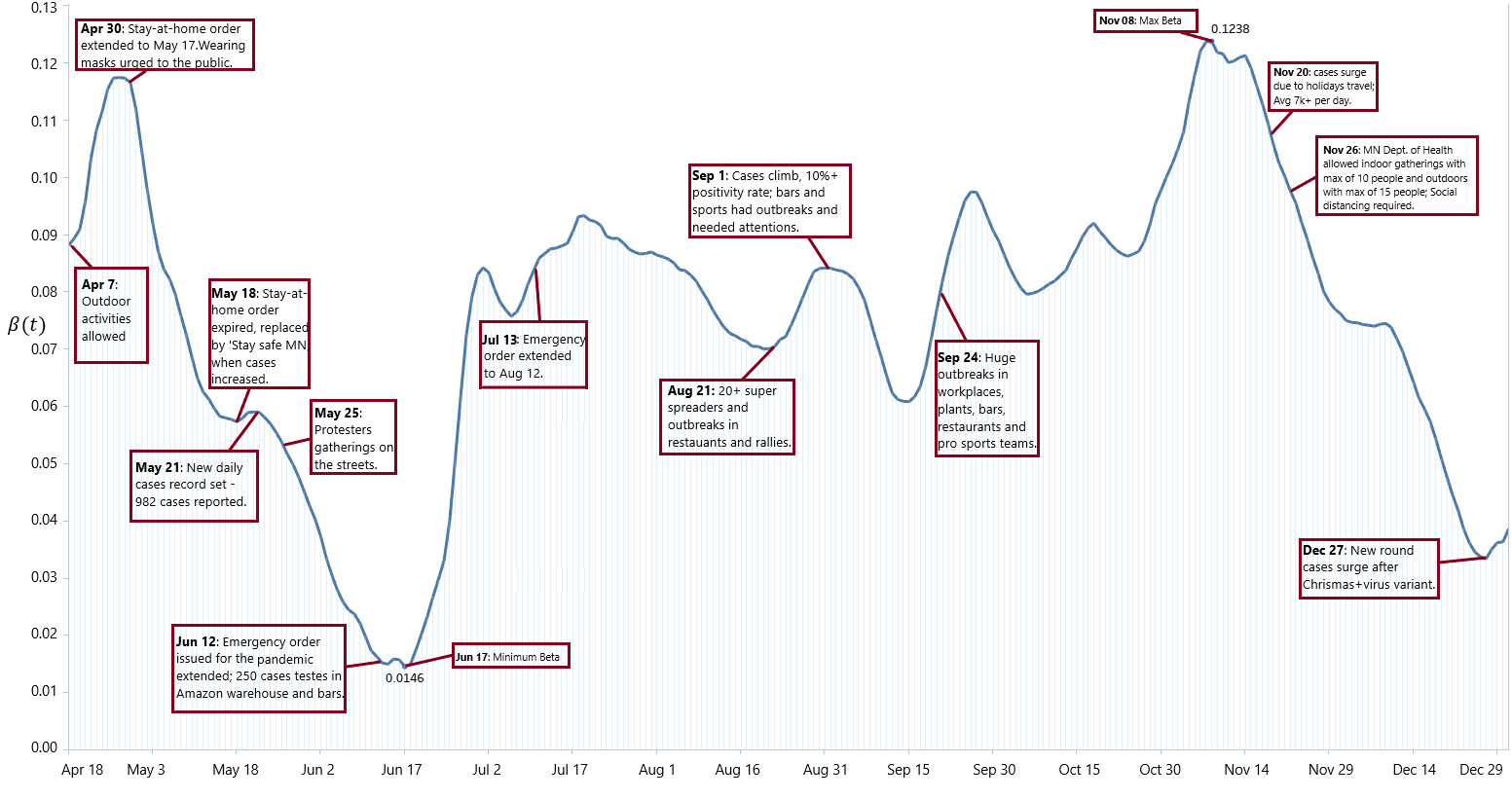}
\caption{Effect of policies against COVID-19 on $\beta(t)$}
\label{fig:beta_MN}
\end{figure}

\begin{figure}[h!]
\centering
\includegraphics[scale=0.45]{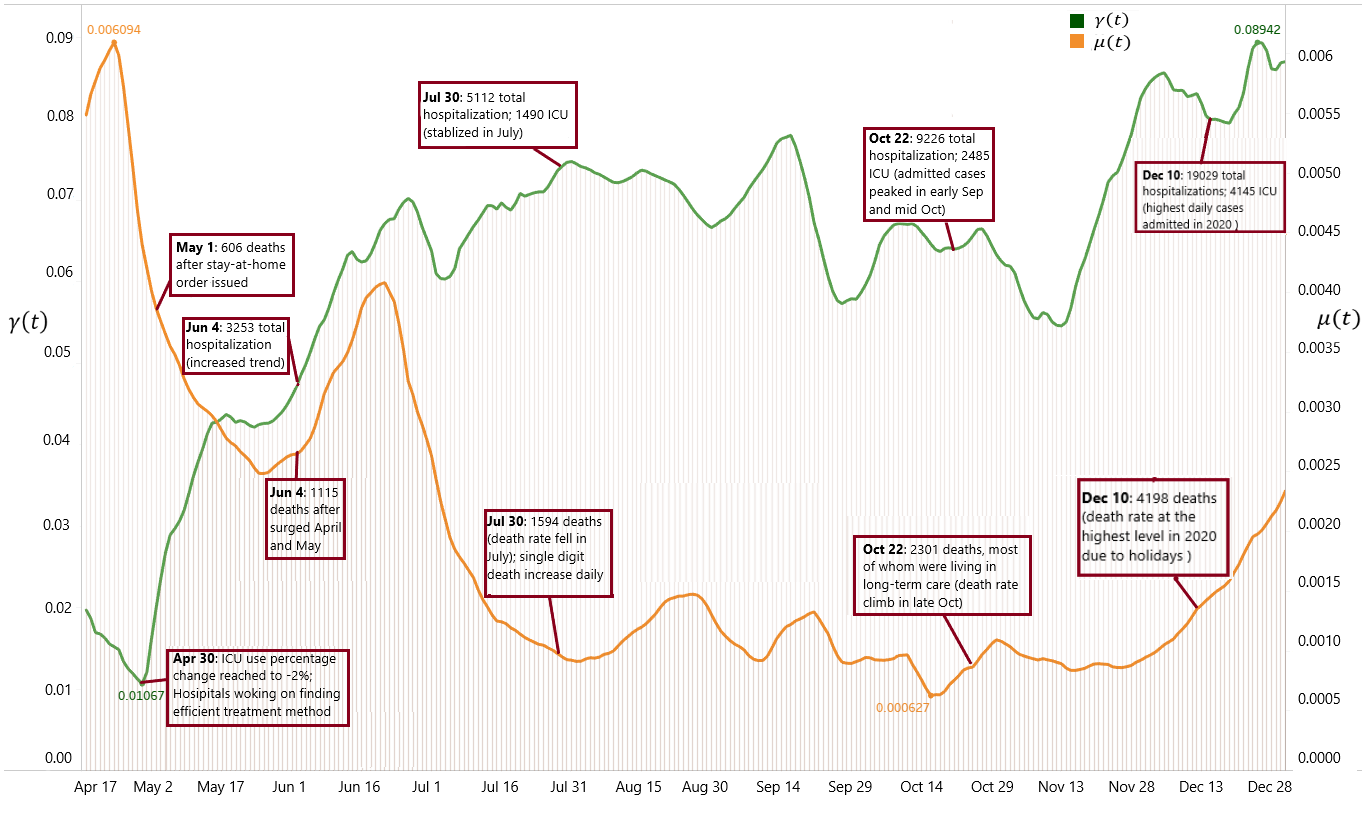}
\caption{Effect of policies against COVID-19 on $\gamma(t)$ and $\mu(t)$}
\label{fig:mu_and_gamma_MN}
\end{figure}

The results obtained for $\beta(t)$ are very sensitive to government policy as well as to people's interaction and observance of social distancing regulations. The time series of $\gamma(t)$ and $\mu(t)$  have a higher dependence on the availability and effectiveness of health care, as well as on the resource availability in hospitals. Figures \ref{fig:beta_MN} and \ref{fig:mu_and_gamma_MN} display timelines of relevant events that took place during the period for which our model found times series of the parameters of the SIRD model. 

A timeline of government policy implementations and social interactions is showed in Figure \ref{fig:beta_MN}. With the first outbreak of COVID-19 in Minnesota, $\beta(t)$ increased  sharply to a peak on April 30\textsuperscript{th}, 2020, when the governor of Minnesota extended the stay-at-home order to May 17, 2020. The order had an effect on the average number of contacts between people and the transmission of COVID-19 slowed down in the first weeks of May. On May 18\textsuperscript{th}, 2020, the new \textit{Stay safe Minnesota} order that contained a series of measures for controlling the pandemic was issued. This new order, along with the effect of masks and social distancing slowed brought $\beta(t)$ to a minimum in the middle of June. However, the impact of these measures backfired shortly after people gathered massively on May 25\textsuperscript{th} to protest against the death of George Floyd. The effect of these demonstrations was a surge in $\beta(t)$ a few weeks later. The emergency order extended until August 12\textsuperscript{th}, 2020 and the higher temperatures during the summer stabilized the situation and even improved it slightly \citep{wang2020high}. However, when the economic activity resumed at the end of the summer, large outbreaks took place in workplaces, bars and restaurants. In late November and early December, when the reported number of new daily cases surged due to holiday gatherings, the state government limited gatherings in an attempt to put a curb on this increase.
 
As mentioned in the quantitative assessment of results, the trends in $\gamma(t)$ and $\mu(t)$ depend largely on the capacity of the health care system to adapt to the hospitalization needs of the population. When hospitals and emergency rooms do not operate at full capacity, they are able to cope better with the needs of their patients and provide better service. This usually results in an increase in the proportion of people that recover (hence, in an increase of $\gamma(t)$) and a decrease in the proportion of people that dies from the disease (hence, a decrease of $\mu(t)$). In the context of COVID-19, the hospitals ability to adapt to the circumstances derived from the pandemic has had a considerable impact since they operated at full capacity during the first months of the pandemic. Moreover, health care professionals have learned more and more about possible treatments to curb on the disease's symptoms and effects. The consequence has been the aforementioned increasing trend in $\gamma(t)$ and the decrease in $\mu(t)$, which can be observed in Figure \ref{fig:mu_and_gamma_MN}. Note that the last weeks of 2020 saw a shift in the trend of $\mu(t)$ as a consequence of the increase of admissions in hospitals following Thanksgiving.

\section{Conclusions}\label{sec:Conclusions}

The present paper provides an alternative framework for describing the time-varying nature of the parameters of an SIRD model. We introduce our work as a ``hybrid" of the works by \cite{anastassopoulou2020data} and \cite{calafiore2020modified}. As in \cite{anastassopoulou2020data}, we propose a rolling regression approach, but we do not set the window size beforehand. As in \cite{calafiore2020modified}, we account for the differences between actual and observed cases, but we allow these differences to vary between infected, recovered, and dead. Our rolling regression is embedded within a MIBNLP problem that allows for selecting an optimal window size for the regression as well as for finding the proportions of the actual number of individuals in each compartment that the collected data represent. In an attempt to reduce the noise inherent to the collection of data, our optimization model looks for a trade-off between the accuracy in reproducing the data that are fed to the model and the smoothness of the time-series of the parameters of the SIRD model. The solution of this MIBNLP problem is tackled with a real-valued genetic algorithm. Using 2020 data from Minnesota, we find that the model yields values for the window size and the fractions of actual cases that our data represent that reproduce satisfactorily the observed daily variations in the number of infected, recovered, and dead individuals. They also provide smooth time series that show more stable and realistic variations in the number of average contacts initiated by infected individuals per unit of time ($\beta(t)$), the proportion of infected people that recover per unit of time ($\gamma(t)$), and the proportion of infected individuals that ultimately die from the disease per unit of time ($\mu(t)$).

We analyze our results under the light of quantitative and qualitative assessments. Quantitatively, our results are in agreement with previously published works, thus proving that our framework is a solid approach that allows for great modeling flexibility. Qualitatively, the time series obtained for the parameters of the SIRD model were consistent with changes in government policy, with people's adoption of social distancing and personal protection measures, and with the improvement of the capacity of the health care system to fight against this new virus. While we apply our research to an SIRD model, we anticipate that this framework can also be used in the context of other compartmental models in epidemiology.

As a future research direction, the implementation of this model opens new questions about the possible relationship between the three parameters of the SIRD models and their variations over time. If there exists an interdependency in these changes and whether that synergy can be captured mathematically is an interesting research question that we would like to address in the near future.







\bibliographystyle{apa}
\bibliography{References_COVID.bib}







\end{document}